\documentclass[prl,byrevtex,balancelastpage,letterpaper,10pt,showpacs,preprint,twocolumn]{revtex4}
\usepackage{amssymb}
\usepackage{amsmath}
\usepackage{graphicx}

\newcommand {\be}{\begin{equation}}
\newcommand {\ee}{\end{equation}}

\begin{document}

\title{Creating ultracold molecules by collisions with ultracold rare gas atoms in an optical trap}

\author{P. Barletta}
\affiliation{Department of Physics and Astronomy, University College London, WC1E 6BT, United Kingdom}
\author{J. Tennyson}
\affiliation{Department of Physics and Astronomy, University College London, WC1E 6BT, United Kingdom}
\author{P. F. Barker}
\affiliation{Department of Physics and Astronomy, University College London, WC1E 6BT, United Kingdom}

\begin{abstract}
 We study collisions of para-H$_2$ with five rare gas atomic species (He, Ne, Ar, Kr and Xe) over the range from 1 K to 1 $\mu$ K and evaluate the feasibility of sympathetic cooling H$_2$ with ultracold ground state rare gas atoms co-trapped within a deep optical trap.  Collision cross-sections over this large temperature range show that all of these species could be used to cool H$_2$ to ultracold temperatures and that argon and helium are the most promising species for future experiments.
\end{abstract}

\maketitle
\preprint{}
\eid{}
\startpage{1}
\endpage{}

The ability to trap and cool dilute atomic gases has revolutionised physics over the last twenty years. Laser cooling has been of primary importance in the exploration of ultracold interactions between trapped atoms, Bose-Einstein condensation, quantum atom optics, and for precision metrology. It has also provided a well-controlled testing ground for condensed matter physics, and more recently quantum information.
Considerable attention has now turned to the creation and study of trapped cold and ultracold molecular gases\cite{revdfk04}. Molecules are much more complex than atoms, offering for instance the possibility of investigating interactions that are not available in atomic species. For example, molecules can have a permanent dipole moment that leads to strong long-range dipolar interactions, while the anisotropic polarizability of molecules offers the ability to orient them in free space and within a trap using electromagnetic fields.  Cold molecules are also seen as an ideal testing ground for physics beyond the standard model of particle physics\cite{coldedm} and as a useful tool in the search for parity violation at the molecular level. The ability to perform chemistry in dilute molecular gases at temperatures far below 1 K promises access to chemical processes dominated by quantum mechanical tunnelling and resonances have yet to
 be observed. \\
In general, laser cooling cannot be applied to molecules due to the absence of a single cycling transition. Although ultracold diatomic molecular species can be produced by association of laser cooled atomic species on Feshbach resonances \cite{revkgj06} and by photoassociation \cite{revjtl06}, the range of molecular species that can be produced is severely limited by the small subset of atomic species that can be laser cooled. New techniques are being developed to produce slow cold molecules of much greater variety and complexity. Many of these techniques, such as Stark deceleration\cite{stark1,stark2}, use conservative fields to filter out a narrow momentum distribution which can be slowed and trapped\cite{electtrap1}. To date all of these approaches been limited to temperatures above 1 mK.  To produce colder temperatures, without significant loss of molecules, it is necessary to employ a dissipative cooling scheme once the molecules are slowed and trapped. A number of important dissipative schemes are potentially available to cool molecules and include stochastic, cavity and evaporative and sympathetic cooling. 
 In sympathetic cooling, the gas to be cooled is co-trapped with a colder gas at higher concentration and heat flows from the hotter gas to the colder gas, coming to a common temperature that is lower than that of the hotter gas. This scheme has been demonstrated experimentally using elastic collisions with cold helium to cool and trap paramagnetic molecules in the 100 mK range, with the ultimate temperature limited by the vapor pressure of helium. More recently however, proposals to use this technique to cool a range of molecules into the ultracold regime (below 1 mK) have been explored using ultracold, laser cooled, alkali species which can be readily created in magneto-optic traps. Theoretical studies to date include collisions of OH and NH with laser cooled Rb\cite{rbohhutson,rbnhhutson}. In these studies co-trapping of the ultracold Rb atoms and molecules in electrostatic and/or magnetic traps is required. Of critical importance to successful sympathetic cooling is the dominance of elastic over inelastic processes over a wide range of collision energies. Inelastic losses that occur due to state changing collisions to untrappable states and/or reactions are of particular concern since these processes lead to losses from the trap preventing thermalization. State changing collisions are particularly important for electrostatic and magnetic traps because the molecules are not usually trapped in their absolute ground state. Finally, the use of an alkali as a collision partner can often lead to losses via reactive collisions. For example, Rb-NH interactions can result in a harpoon reaction which creates the RbNH molecule\cite{rbnhhutson}.

We propose a different approach to sympathetic cooling of molecules that eliminates both of these loss mechanisms and additionally can be used with a range of deceleration and trapping methods which have been demonstrated to produce stationary, stable and cold molecules below 1 K. This method uses inert, rare gas atoms as the ultracold collision partner and co-traps the molecules and atoms in their absolute ground state in a deep optical trap. Ground state rare gas atoms cannot be directly laser cooled, but when promoted to their metastable triplet state via electron impact they can be laser cooled and trapped in a magneto-optical trap (MOT)\cite{hecool,arkrcool,necool,xecool}. Once cooled they can be quenched to their ground state via excitation to a higher lying state which spontaneously decays to the absolute ground state\cite{quench}. Cold decelerated molecules can then be spatially overlapped with the cloud of ground state rare gas atoms in a deep quasi-electrostatic optical trap (QUEST) or a microwave trap and both species can be held simultaneously.  In these traps all states are high field seeking so even state changing collisions do not prevent trapping and thermalization. However, as relatively warm Stark decelerated molecules in the 10-100 mK range must be trapped as well, high intensity fields must be used to create a deep trap.

 A buildup cavity offers a route to creating high intensities and the well depth from such a QUEST is given by 
\be U = \frac{U_0}{1+(x/x_r)^2} \exp[-2(r/\omega)^2] \cos^2[kx],\ee
 where the well depth is given by $U_0 = \frac{2\alpha}{\varepsilon_0 c}I_c$, and $\alpha$ is the static polarizability, $I_c$ is the one way circulating peak intensity and $k = 2\pi/\lambda$ is the wavevector along the axis of the trap taken to be along $x$.  The $1/e^{2}$ width of the field in the radial direction $r$ is given by $\omega$ and the Rayleigh range of the intracavity field is $x_r = \pi\omega^2 / \lambda$. The radial and axial oscillation frequencies for particles in the trap are given respectively by $\omega_r = [4 U_0/m \omega^2]^{1/2}$ and $\omega_x = [2 U_0 k^2/m]^{1/2}$. Intensities in the 10$^8$ W cm$^{-2}$ range can be produced in an optical buildup cavity using laser input powers of 17.5 W and a cavity finesse of 6300 \cite{cavitydamage}.  This intracavity intensity corresponds to a well depth of 70 mK for C$_6$H$_6$, 19 mK for NH$_{3}$ and 6 mK for H$_2$.  For the ground state rare gases, depths of 27 mK (Xe), 17 mK (Kr), 11 mK (Ar), 3 mK (Ne), and 1.4 mK (He) could be achieved using this design and it is feasible that much deeper optical traps could be created using a near concentric cavity, but with a reduced trapping volume. The well depths achieved for the larger more polarizable molecules compare favourably with that achieved to date using electrostatic traps for molecules and could be combined with these traps to increase phase space density. The deeper well depths also lead to higher oscillation frequencies which will reduce the time to thermalize species withi the traps. As rare gas atoms in their ground state have a relatively high ionization potential (lowest is 12.1 eV for Xe) they are robust against multiphoton/tunnel ionization within the high intensity fields of the trap. Similarly, most stable molecules of interest also have high ionization potentials and no ionization is seen using intensities even in the 10$^{12}$ Wcm$^{-2}$ range \cite{opticalstark1}.  The choice of which rare gas species to use for sympathetic cooling must depend primarily on favorable elastic to inelastic collision cross sections between the species, but also on the polarizability of both the atom and molecular species. Of the rare gas atoms considered, xenon is the most polarizable and the deepest traps can be created for this species. However, the lighter rare gases such as helium and neon may be more favorable for cooling lighter molecules due to the favourable mass ratio and weaker Van der Waals interactions.

The aim is develop a method for trapping arbitrary, stable (ie closed shell)
molecules.
To explore the feasibility of this scheme we have chosen to study
collisions of H$_2$ with ultracold rare gases. Study of these
collisions is important for their fundamental nature, but also for
their importance in testing {\it ab initio} theoretical models and their
relevance for astrophysics. H$_2$ is also seen as a ideal candidate to
study chemistry at ultracold temperatures dominated by resonance and
tunneling phenomena \cite{h2fchem,h2fexp}. Cold H$_2$ in the 10 mK to
100 mK range can in principle be decelerated using optical Stark
deceleration \cite{opticalstark1, opticalstark2} from an intense
molecular beam which can be created in pure para form with initial
temperatures in the 100 mK range.

 Cold and ultracold collisions and calculations of collision cross sections for H$_2$ with He and Ar have been already considered in some detail\cite{arh21,hemcf06}. In one sense, we extend these to include all stable rare gas atoms that can be laser cooled and trapped. On the other hand, much of the previous work focused on internally highly excited H$_2$ molecules, and was directed to study rovibrational energy transfers. For this purpose, a system of coupled equation needs to be solved containing at least as many equations as the number of open channels \cite{hemcf06}. The experimental setup proposed aims at producing H$_2$ that is decelerated for example by optical Stark deceleration in its vibrational-rotational ground state. Under this very specific experimental condition, the scattering of Rg-H$_2$ can be effectively be modelled as a two-body process \cite{heggl05}. As the scattering energy is very small compared to the vibrational-rotational excitation energies of H$_2$, the only allowed channel is the elastic one, and no inelastic processes need to be considered. Under those assumptions, the Rg-H$_2$ system can be effectively modelled as a two-body system \cite{pbxiv08}, with an effective Rg-H$_2$ interaction
\be
V_{eff}(R)=\int \phi^2_{00}(r) P^2_0(\mu) V_{\rm Rg-H_2}(r,R,\mu) dr d\mu ,
\ee
where $r$ is the H-H distance, $R$ is the Rg-H$_2$ dissociation coordinate, $\mu=\hat{r}\cdot\hat{R}$ and $P_0$ is a Legendre polynomial. $\phi_{00}$ represents the ground vibrational-rotational function of H$_2$, and $V_{\rm Rg-H_2}(r,R,\mu)$ is the RgH$_2$ potential energy surface. From the asymptotic behaviour of the Rg-H$_2$ wavefunction at large $R$ one can extract the phase-shift $\delta(E)$, thus obtaining the elastic cross-section $\sigma(E)$. The zero energy limit of $\sigma(E)$ is well known, and for $s-$wave scattering is $\sigma(E=0) = 4 \pi a_s^2$, where $a_s$ is the Rg-H$_2$ scattering length. Therefore the determination of the scattering length provides a functional tool to compare the ultracold scattering properties of the different complexes. 

The scattering length can be dramatically sensitive to small variations in the potential energy surface (PES). This is particularly true in systems which are very weakly bound. Therefore, the accuracy of the underlying molecular PES is essential for a sensible determination of the Rg-H$_2$ $a_s$.  As a consequence of a long standing interest in Rg-H$_2$ complexes, the Rg-H$_2$ interaction has been studied in great detail, in particular for ArH$_2$ and HeH$_2$. However, the Rg-H$_2$ Van der Waals interaction is weak due the closed shell structure of both H$_2$ and the rare gas atoms. Despite many years of work, there is still a small but critical uncertainty associated with the Rg-H$_2$ PES's.  For example, in HeH$_2$, which is the weakest bound of all five complexes, the existence of a bound state has been long debated, and only recently the calculated potentials have become precise enough to confirm its existence \cite{hekkr04}.  To estimate the scattering lengths of the various complexes, we have gathered several available PESs from the literature. Table \ref{tab1} reports the range of values obtained for the Rg-H$_2$ scattering length and effective range. Of all complexes and isotopologues, $^3$HeH$_2$ represents a special case as the percentage variation in its mass makes its collisional properties sufficiently different from  $^4$HeH$_2$ to make it worth mentioning explicitly. The values of table \ref{tab1} have been calculated using a single channel trial wavefunction, and corrections were also calculated and proved insignificant \cite{pbxiv08}. Balakrishnan {\it et al.} \cite{hebfd98,helrm05} performed a detailed calculation of the He-H$_2$ collisional process using both PES's of Ref. \cite{hemr94} and Ref. \cite{hebmp03}; our results agree well with theirs.  Table \ref{tab1} shows that, although there are big uncertainties in the collisional parameters associated with the uncertainties in the PES, the physics of the scattering mechanisms is rather clear. The greatest cross-section is of He-H$_2$, due to the strong halo characteristics of this system \cite{hekkr04,heggl05}. The concept of quantum halo systems is borrowed from nuclear physics \cite{revjrf04} and defines quantum systems which extend well into the classically forbidden region. It is possible to show that for those systems the scattering length depends on the square root of the inverse of the binding energy E, $a_s \approx 1/\sqrt{E}$, therefore the smaller the binding energy the larger the scattering length. The scattering lengths of the five complexes can be interpreted in terms of this relationship. For example, the two HeH$_2$ isotopologues are by far the weakest bound of the series thus their scattering length is the greatest. The NeH$_2$ system is more strongly bound, therefore its scattering length is smaller. The stabilization of a  second vibrational band in ArH$_2$ makes this complex more reactive than its predecessor. Again, as the potential binding increases, the KrH$_2$ $a_s$ is smaller than the one of ArH$_2$. The Xe-H$_2$ scattering length is the smallest of the whole group.


\begin{figure}[h]
\includegraphics[scale=0.35,angle=-90]{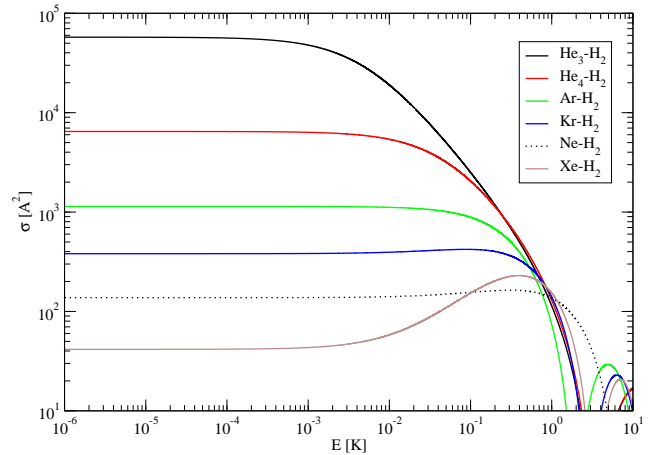}
\caption{Collisional cross-sections for Rg-H$_2$ complexes over the energy range from 10 K to 1$\mu$ K}
\label{fig1}
\end{figure}

\begin{table}[h]
\begin{center}
\begin{tabular}{c|rrrr} \hline \hline
              &  $|a_s|$ [\AA] & r$_{eff}$ [\AA] & $\sigma(E=0)$ [\AA$^2$] & PES Refs.  \\ \hline
 $^3$He-H$_2$ & 67.6-90.6    & 6.08-6.27     & 57500-103000            & \cite{nersc82, hemr94, hebmp03} \\
 $^4$He-H$_2$ & 22.7-24.7    &5.00-5.16      & 6500-7800	       & \cite{nersc82, hemr94, hebmp03} \\
     Ne-H$_2$ & 3.30-3.85    &10.7-14.9      & 140-190 		       & \cite{neabh80, neh21, nersc82} \\
     Ar-H$_2$ & 8.71-10.1    &5.21-6.12	     & 950-1300		       & \cite{nersc82, xerh86, arh22, arh25} \\
     Kr-H$_2$ & 5.51-6.96    &9.65-14.8      & 380-610		       & \cite{xerh86, krh21, krh22}  \\
     Xe-H$_2$ &  1.82        & 289          &  42                      & \cite{xerh86} \\ \hline \hline
\end{tabular}
\caption{Ultracold collisional properties for Rg-H$_2$ complexes, namely the scattering length a$_a$ and the effective range r$_{eff}$. The zero-energy elastic cross section $\sigma$ is also tabulated.}
\label{tab1}
\end{center}
\end{table}

The XeH$_2$ PES is the one with the greatest uncertainty. Tests on the
sensitivity of the results of Table \ref{tab1} to the potential showed
that Ne-H$_2$ and Kr-H$_2$ are essentially unaffected by increases or
decreases of the interaction potential by over 25~\%. However the
$a_s$ for Ar-H$_2$ and Xe-H$_2$ can become comparable to the one of
$^4$He-H$_2$ with  relatively small adjustment of their PES, $-25$~\%\ and +
25~\%\ respectively. As the Ar-H$_2$ PES is weakened, the already
weakly bound excited vibrational band becomes closer to the
dissociation threshold, whereas for the Xe-H$_2$ strengthening the
potential brings a third vibrational state close to being bound.


The large collisional cross-sections for $^4$He-H$_2$ and Ar-H$_2$ are comparable to that utilized in sympathetic cooling experiments between cold alkalis\cite{licssymp1}and these values are relatively constant over a large temperature range.  We estimate a collision rate and thermalisation time for Ar- H$_{2}$ based on a cross-section of 1000 {\AA}$^2$  and radial and axial trap frequencies of 44 2$\pi$ kHz and 37.4 2$\pi$ MHz for Ar and 16.5 2$\pi$ kHz and 13.8 2$\pi$ MHz for H$_2$.  We also assume that H$_2$ has evaporated to approximately 1/10 of the well depth before sympathetic cooling occurs. We assume that each species has a density in the trap of the form $n=n_{0} \exp{[-U/kT]}$ with peak densities of $5\times10^{9}$ cm$^{-3}$ and $10^{11}$ cm$^{-3}$ corresponding to 11 molecules and 1300 atoms per fringe of the trap at initial temperatures of 600 $\mu$K and 10$\mu$K respectively. 
The Rg-H$_2$ collision rate is given by
$\gamma = \sigma  v_{rel} \int n_{\textrm{Rg}}(\bold{x}) n_{\textrm{H}_{2}}(\bold{x}) d\bold{x}$,
  $v_{rel}$ is the mean relative velocity between collision partners and $n_{Rg}(\bold{x})$ and $n_{H_{2}}(\bold{x})$ are the densities of the rare gas and $H_2$.  It is well known that approximately 3 collisions are required for thermalization of collision partners of equal mass in a gas but for unequal masses this is approximated by $\frac{3}{\eta}$ where $\eta = 4 \frac{m_{\textrm{Rg}} m_{\textrm{H}_{2}}}{(m_{\textrm{Rg}}+m_{\textrm{H}_{2}})^{2}}$. The initial time for thermalization for two species within a trap is given by \cite{licssymp1}
\be \tau= \frac{3\pi^2 k_b T_{{\textrm H}_2}}{(N_{{\textrm H}_2} + N_{\textrm{Rg}})\sigma \eta m_{{\textrm H}_2}{\omega_{z}}_{{\textrm H}_2}{\omega_{r}}_{{\textrm H}_2}^2 }. \ee 
 For the conditions given for Ar-H$_{2}$, a thermalization time scale of 300 ms is determined. This relatively fast thermalization time can be accounted for by the large axial trap frequency and the large relative velocity (2.5 ms$^{-1}$) between the hot molecular gas and the much colder atoms, despite the poor spatial overlap between the two species.  For Xe-H$_{2}$ collisions, using the same initial densities and a collision cross-section of 42 {\AA}$^2$, a thermalization time of approximately 30 seconds is estimated. Although the Xe-H$_{2}$ collision rate is much lower than for Ar-H$_{2}$ trapping of atomic species over this time scale has been demonstrated indicating that even Xe, which forms the deepest trap, could be used for sympathic cooling.

We present a new route towards the creation of ultracold molecules by sympathetic cooling with laser cooled metastable rare gas atoms that are quenched and subsequently co-trapped in a deep optical lattice trap. The scheme allows the simultaneous trapping of relatively 'warm' molecules and ultracold rare gas atoms and utilizes the high trap frequencies in deep standing wave traps for increasing the collision rate and reducing thermalization times.  In order to evaluate this scheme for cooling molecular hydrogen we have calculated the collisional properties for all five rare atoms. Using the cross-sections determined in this study we show that sympathetic cooling of H$_2$ molecules by all five atoms is technically feasible over a wide temperature range from the cold to ultracold regime and that this method may be promising for a large range  of molecular species.

\end{document}